# Data Want to be Free: An Innovation Resistance Theory Model for Identifying Barriers to Government Data Sharing

Anastasija Nikiforova[1] [1][0000-0002-0532-3488], Antoine Clarinval[2][0000-0003-3936-0532], Anneke Zuiderwijk - van Eijk[3][0000-0002-3552-7289], Daniel Rudmark[4][0000-0003-3554-2267], Petar Milic[5][0000-0003-0427-8379], Charalampos Alexopoulos[6][0000-0002-6610-0675], Katrin Rajamäe-Soosaar[7][0009-0000-2637-8202]

[1] University of Tartu, Estonia
[2] Umeå University, Sweden
[3] Delft University of Technology, the Netherlands
[4] The Swedish National Road and Transport Research Institute, Sweden
[5] University of Priština - Kosovska Mitrovica, Serbia
[6] University of the Aegean, Greece
[7] Association of Estonian Cities and Rural Municipalities, Estonia

**Abstract.** *Data sharing is increasingly essential for digital government and data-driven innovation, yet many public organizations remain reluctant to make their data openly available. While prior research has examined factors influencing open data adoption, little theoretical work explores why resistance persists within public agencies. This study develops an Innovation Resistance Theory (IRT) model tailored to government data sharing to identify predictors of organizational resistance. An initial model was derived from literature and refined through interviews with 21 public organizations across six European countries. The resulting IRT4DS model identifies 39 barriers spanning usage, value, risk, tradition, and image dimensions, and 23 countermeasures mapped to the most critical barriers and the actors responsible for addressing them. By extending IRT into the context of governmental data sharing, the study advances theoretical understanding of why public data often remains closed and provides actionable guidance for policymakers seeking to design enabling data ecosystems and reduce structural and cultural barriers to OGD adoption.*

[1] Corresponding author: Anatsasija Nikiforova, Anastasija.Nikiforova@ut.ee, Nikiforova.Anastasija@gmail.com

# 1 Introduction

Echoing the well-known maxim that "*information wants to be free*,"[2] government data are widely viewed as a foundational public resource. Open Government Data (OGD) have emerged as a catalyst for sustainability-oriented and inclusive data-driven innovation, enabling value creation for citizens, businesses, researchers, and public organizations [17,46,61]. As open data ecosystems evolve from basic transparency towards strategic and systemic reuse, OGD increasingly form the backbone of digital innovation. Companies develop new services using OGD [51], researchers combine datasets for scientific discovery and modelling, including crisis management [57], public agencies use each other's data to improve collaboration and service delivery [52]. Yet, despite their recognized value, many public organizations continue to resist (actively or passively) opening their data.

Prior research on OGD sharing has predominantly applied acceptance-oriented adoption theories such as Technology Acceptance Model (TAM) [56] and the Unified Theory of Acceptance and Use of Technology (UTAUT) [57,62]. While these approaches illuminate motivational drivers, they pay limited attention to the *barriers* inhibiting openness. Existing studies identify issues such as low data quality, lack of technical skills, resource constraints, fears of misuse [2,15,52], yet insights remain fragmented. At the same time, e-government scholars increasingly call for renewal beyond repeatedly applied adoption models that offer diminishing explanatory value [1,21]. This study fills these gaps by examining barriers to OGD sharing through Innovation Resistance Theory (IRT). IRT offers a theoretically rich framework by conceptualizing resistance as stemming from (a) functional barriers - usage difficulties, value of innovation, risk perceptions, and (b) psychological barriers - cultural misalignment, image concerns [44]. Viewing OGD as an innovation adopted by public agencies [17], IRT enables the integration of diverse barriers into a coherent explanation of resistance. Although widely applied in marketing and consumer innovation [12,27], it remains underutilized in e-government.

The objective of this study is to develop and validate an IRT-based model tailored to OGD—IRT4DS (Innovation Resistance Theory for Data Sharing)—that identifies predictors of resistance among public agencies. We develop an initial model from literature and refine it through interviews with 21 agencies across six countries. The final model comprises 39 barriers across the five IRT categories and is complemented by 23 countermeasures mapped to the most relevant barriers. This theoretically grounded and practically actionable framework contributes to IS and digital government scholarships by serving as a diagnostic model explaining the mechanisms through which data becomes open or remains enclosed and offers insights essential for building resilient open data ecosystems and supporting digital transformation in the public sector.

Our empirical findings confirm known barriers but also reveal several underexplored or novel barriers, such as *format-specific capability gaps* (e.g., difficulty publishing in required GIS or RDF formats), *capacity mismatch* due to the absence of OGD-dedicated roles, *historical path dependence*, where openness conflicts with inherited mandates and routines, the *ongoing maintenance burden, financial cannibalization risks* for agencies with legacy revenue models, and *difficulties prioritizing* under limited capacity. By surfacing these mechanisms, the study advances theoretical understanding of why public data remains closed and demonstrates the analytical value of IRT for OGD research, revealing barriers that have remained theoretically insufficiently visible but have practical implications.

[2] expression that means that information naturally strives to become as freely available among people as possible (attributed to Stewart Brand, 1984 quote), later transformed to denote that all people should be able to access information freely

## 2 Background and Related Work

### 2.1 Innovation Resistance Theory

Resistance to change has long been viewed as a natural response when established routines are challenged [43,59]. Innovation resistance, in turn, was defined in [44] as a special version of resistance to change widely discussed in social psychology and behavioural science [18,36] sharing a common opinion that individuals have an inherent / intrinsic desire for psychological balance called "*psychological equilibrium*." This is because any change made to the individual's behaviour can disturb this balance, so the individuals often choose to resist change rather than go through the most likely disturbing and challenging process of readjustment. Therefore, it is believed that resistance is a normal response of individuals when faced with innovation [43]. IRT argues that adoption outcomes reflect the interplay of *functional* and *psychological* barriers. Functional barriers relate to *usage*, *value*, and *risk* perceptions, while psychological barriers derive from *traditions* and norms of individuals, and *image* concerns. Together, they explain why innovations are *rejected*, *adopted*, or selectively *modified* [43].

Although the IRT was developed over 30 years ago, it remains popular and has gained rapidly increasing popularity in the last 5 years. Most scholars used the IRT as the basis for the empirical evaluation of consumer resistance to innovations. Our literature analysis[3] suggests that existing research utilizing IRT can be classified into 5 high-level groups depending on their application domain: (1) digital financial services, e-commerce and online services (Talwar et al., 2020); (2) sustainable consumption, green innovations and organic food [55]; (3) cutting-edge technologies and emerging trends [12,27]; (4) education and MOOC [32]; (5) e-government. We identified only two studies utilizing IRT in e-government. Prakash and Das [42] explored citizens' resistance to use digital contact tracing apps that governments worldwide used as a critical element in their COVID-19 pandemic lockdown exit strategy. Nikiforova and Zuiderwijk [38] presented the initial version of the IRT model tailored to OGD that we validate and extend in this study. IRT remains underutilized in e-government research, where its application could help uncover organizational, institutional, and operational mechanisms of resistance overlooked by traditional acceptance-oriented adoption theories. Viewing OGD as an innovation adopted by public agencies [17], IRT enables the integration of diverse barriers into a coherent explanation of resistance. Although IRT was originally developed to explain consumer resistance to innovations, its core mechanisms are equally relevant in organizational contexts where innovation adoption decisions are shaped by institutional routines, professional norms, and reputational considerations. In public sector organizations, these mechanisms often manifest collectively through managerial decisions, regulatory constraints, and organizational culture.

### 2.2 Barriers to Openly Sharing Government Data

Research on OGD consistently identifies barriers that impede data opening, ranging from technical constraints to institutional and cultural factors [2,46]. Synthesizing prior work through the lens of IRT shows how these barriers map to its functional and psychological categories (Table 1): (1) usage barriers refer to data quality problems, complex publication processes, and low usability of portals or data formats; (2) value barriers - to unclear benefits, resource requirements, incomplete datasets, and concerns over lost revenue streams; (3) risk barriers involve fears of data misuse, misinterpretation, exposure of sensitive information, legal liability; (4) tradition barriers include risk-averse cultures, rigid routines, and limited skills; (5) image barriers stem from concerns that published data may lead to reputational harm if misused or if quality is questioned.

[3] See full analysis on ZENODO - https://doi.org/10.5281/zenodo.15291553

**Table 1.** Examples of barriers to sharing OGD (revisited [38]).

| Resistance factors | Examples of barriers to openly sharing OGD leading to resistance |
|---|---|
| Usage barriers | OGD often suffer from quality issues [48;57]<br>Openly sharing government data is a complicated process [2]<br>OGD portals suffer from low ease of use [7;14]<br>Insufficient user-friendliness of the data [33] |
| Value barriers | OGD do not always provide value to users [10;29;53]<br>Datasets may be incomplete [37;63]<br>Concerns about the quality of open data [3;33;37;57]<br>Sharing government data requires resources, including time and costs [52]<br>Impossible to sell the data when it is openly available [33;9]<br>Data providers are usually the ones who invest the most effort in publishing data, while businesses and citizens profit the most [53;29] |
| Risk barriers | Fear that openly shared government data will be misused [53]<br>Fear of open data users drawing false conclusions [2;9;57]<br>Fear that (privacy) sensitive data will be shared openly [2;33;46;48]<br>Fear of making mistakes when preparing data for publication [26;63]<br>Fear of being liable for data quality [2] |
| Tradition barriers | Risk-averse culture avoids openly sharing the data [56]<br>Reluctance to change organizational processes [9;52]<br>Incompatible organizational routines and processes [31]<br>Civil servants may lack the skills required for openly sharing data [29;31;52] |
| Image barriers | Fear that organizational reputation will be damaged due to the publication of low-quality data [7;11;63]<br>Fear that the organization sharing the data will be associated with incorrect conclusions drawn from OGD analysis / misinterpreted [15;31] |

While these studies offer valuable insights, they typically analyze individual barriers or isolated categories. What is missing is a coherent, theoretically grounded model integrating functional and psychological resistance factors. IRT provides such a structure, allowing diverse barriers to be systematically examined as part of a unified resistance mechanism rather than as disconnected obstacles.

## 3 Methodology

To develop an IRT-based model tailored to OGD (IRT4DS), we first conducted a literature review on both IRT and barriers to data sharing and used these insights to construct an initial model that has been outlined in [38] (summarized in Section 2). This model is now validated and refined through qualitative interviews with public organizations. Qualitative inquiry is suitable for uncovering context-specific mechanisms, identifying unforeseen factors, refining theoretical constructs, thereby contributing to the theory building [6].

### 3.1 Data Collection

We interviewed 21 public agencies across six European countries selected to reflect different administrative traditions [40], geographical areas, and unitary and federal states, and levels of open data maturity according to the Open Data Maturity Report 2022 (the most recent report at the time of sampling countries). Considering that IRT supposes consideration of non-adopters of innovation as those that potentially face actual barriers associated with adoption, we were interested in followers and beginners. The sample includes Latvia, Belgium, Greece, the Netherlands, Sweden, and Serbia. Following [5], a cross-country case study methodology has been developed to dive deeper into individual cases [34,58] without losing sight of the bigger picture [28].

Within each country, we used purposive sampling to recruit agencies with differing mandates and roles [41]. We selected public agencies based on their availability for interviews and aimed for diversity in the type of organization (e.g., thematic public service, statistical institute, etc.). A total of 21 interviews (i.e., one per public agency) were conducted with 25 interviewees, including directors, project managers, advisors, and data specialists. 17 interviews involved 1 respondent and 4 were done with 2 respondents. Although country-level comparisons were not the objective of this study; instead, data were analysed at the aggregate level, with a comparable number of interviews in each country. Interviews were conducted between May 2022 and April 2023. Table 2 shows the distribution of interviewees per role.

**Table 2.** Distribution of interviewees per role.

| Role | Number of interviewees | Interview ID |
|---|---|---|
| Project manager | 8 | I1, I4, I5, I8, I11, I12, I13, I18 |
| Director | 7 | I2, I3, I9, I10, I17, I20, I21 |
| Manager | 4 | I6, I15, I16 |
| Advisor | 4 | I2, I7, I14, I19 |
| Data analyst or scientist / researcher | 2 | I3, I5 |
| **Total** | **25** | **-** |

Interviews followed a semi-structured protocol derived from the initial model and its individual barriers (i.e., items)[4]. To ensure the reliability and validity, the items selected as constructs in our model were pre-tested by two researchers. Before the interviews took place, the items were checked for content validity and unambiguous wording. Protocol includes two major sections. The first section consists of 9 questions about the general profile of the organization that the respondent represents and their experience with OGD. The second section contains IRT model-related questions. Considering the explorative nature of the study and our intention to refine the model, each individual barrier was transformed into an open-ended question of the form "*To what extent do the following situations form a barrier to openly sharing your organization's data: …*" where each barrier was then addressed. Open-ended questions were given preference over closed-ended questions (yes/no) due to the exploratory nature of our study. Before the specific barriers within a category are addressed, the respondent is introduced with a definition of the barrier category (i.e., resistance sub-factor) and asked whether there is any situation that would pose a challenge in the scope of the category for the organization. Then, after going through each specific barrier within the category, the respondent is asked whether they can think of any other relevant barrier. This is repeated for the five barrier categories. Once all the barriers and their associated questions are asked, another general question enquires about other barriers not mandatorily related to the above categories. To avoid language barriers, the interview protocol was translated into the respondent's language. Before the interview, the purpose of the study was introduced, and informed consent was obtained for recording and further use of the collected data. Afterward, respondents were asked to confirm the transcription accuracy.

[4] Available in full at https://doi.org/10.5281/zenodo.15291553

### 3.2 Data Analysis

The thematic analysis of the collected data and interpretation of the responses was carried out according to [13] and [39]. Interview transcripts were reviewed and coded accordingly by two researchers to whether each barrier was perceived relevant, partly relevant, or irrelevant in the interviews. Coding discrepancies were discussed and resolved through iterative consensus-building sessions involving the full research team. Each barrier was ranked from low to very high (low/moderate/high/very high). If the number of respondents who consider the barrier relevant or partly relevant for them is below or equal to 10%, we consider the relevance of the individual barrier to be *low*, between 11 and 20% - *moderate*, between 21 and 50% - *high*, and over 50% - *very high*. Based on this, we formulated propositions to the initial model, such as reformulating, merging, adding, and removing barriers and reached consensus among the authors team (seven researchers in total) and refined the model accordingly.

## 4 IRT4DS - IRT Model Tailored to OGD

### 4.1 Experience with OGD and Barriers to Sharing

The interviewees demonstrated diverse experiences with data sharing, driven by legal obligations, mission mandates, internal efficiency goals, or societal expectations. All respondents reported challenges related to technical issues, legislative ambiguity, limited understanding of available data, or the absence of a formal mandate. Staffing and infrastructure constraints further complicate publication. Despite these obstacles, there is a growing acknowledgment of the importance of OGD and expressed efforts to improve their practices. Frequent mentions of resistance to change and innovation (I5, I7, I11, I12, I14-I19, I21) during the interviews support the relevance of IRT for explaining their experiences. Below, we summarize the insights obtained in the interviews across the five IRT barrier types, focusing on 10 barriers considered relevant by the most interviewees, complemented with most relevant (individual) insights. Figure 1 shows the distribution of relevance ratings, and Table 3 provides the full set of refined barriers.

**Usage Barriers.** Twelve (12) out of 21 respondents noted difficulties preparing data in line with OGD principles (UB2), although, as per I2, "*the complexity does not mean that the government does not open the data*." The process tends to be recognized as challenging due to limited experience (I2, I8), organizational size and expertise (i.e., lack of expertise within smaller organizations) (I3), shortages of human resources (I3, I16, I21), and the technical demands of certain data types, i.e., opening geospatial data can require programming and additional security and data protection measures such as masking and anonymization (I14, I18).

Process automation (UB7) was frequently mentioned as desired but rarely available on portals (I1, I9, I14, I16, I17, I18). To this end, some organizations developed internal scripts for semi-automation (I3, I5). Although automation is often perceived as a technical barrier, the lack of a dedicated system to manage open data processes was found as not merely a functional barrier, but rather a cultural and motivational barrier (I14). Furthermore, I6 noted that they do not fully automate the process -although they could- because publication would outpace the data validation process. Respondent I5 also mentioned the *lack of willingness (low sense of priority) to open data*, and the *difficulty of the data prioritization process*, especially when an organization deals with large volumes of data, different data providers from whom data can be obtained, or data providers collecting similar data with implied difficulty to select the most appropriate provider with whom to collaborate, highlighting the strategic nature of this barrier.

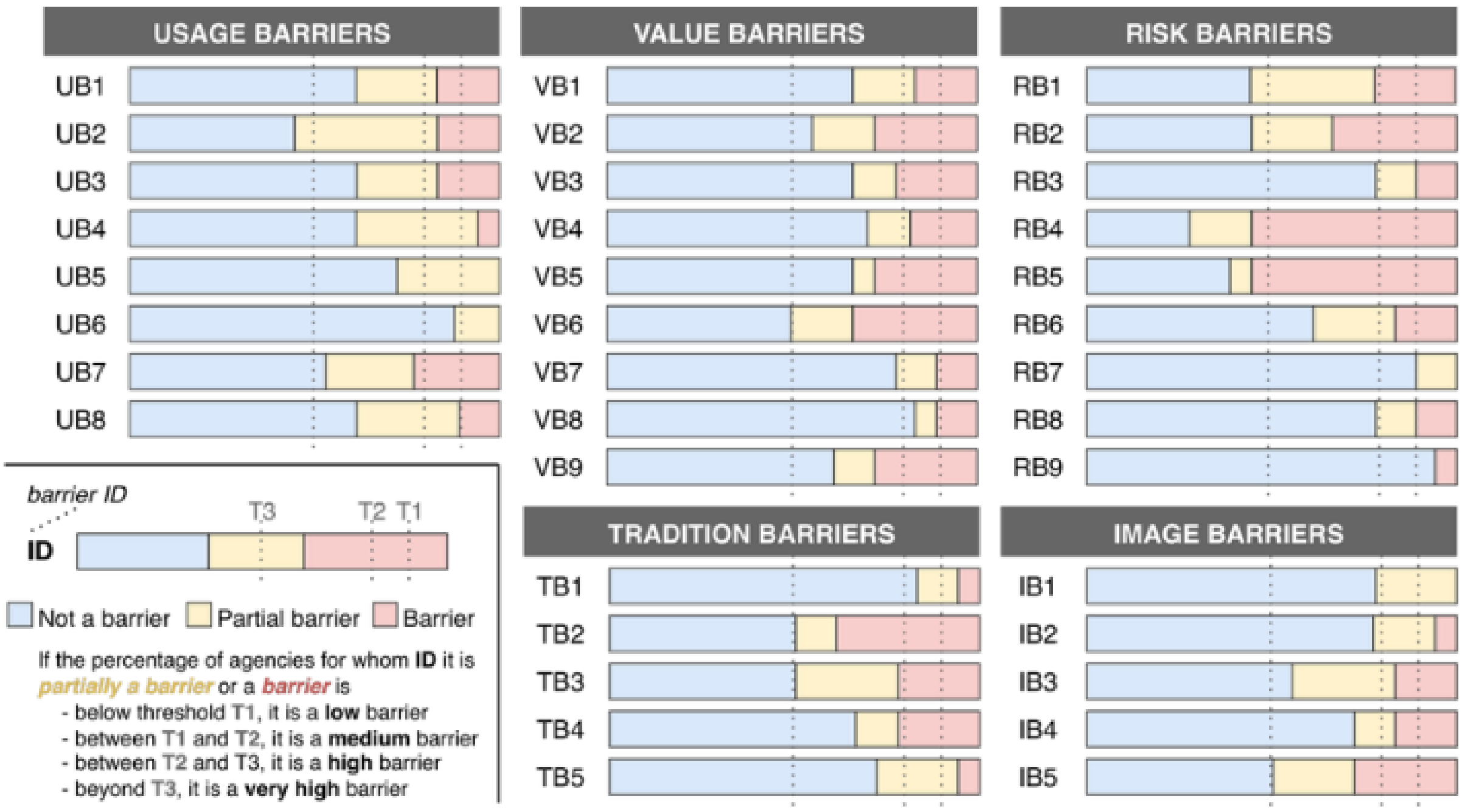

**Figure 1.** Relevance assessment of the barriers by interviewees (for list of individual barriers, see Table 3).

**Value Barriers.** Eleven (11) respondents consider tasks associated with data preparation to be resource-consuming (VB6), with ten respondents, however, not considering this a major barrier that will prevent them from publishing data, rather mentioning delays in publication, manual processes, or legal obligations, manageable with the right means or expertise. Some respondents consider data preparation to be a major obstacle due to the substantial financial or staffing pressures (I4, I8, I10, I16, I17, I18). Over time these burdens were sometimes mitigated by process redesign (I13), accumulated experience, or better understanding of OGD's value.

**Risk Barriers.** Ten (10) respondents expressed concerns about data misuse (RB1), especially when data released in machine-readable formats could be exploited more easily than non-open formats - "*PDF is not dangerous, but an Excel table is*" (I5). This fear was particularly strong among agencies managing geospatial or national security-sensitive information, since, according to I7, "*you need a good map to win a war*." I1, for whom this was a barrier at the beginning of data opening, explained that there was a lot of discussion about whether it is possible to open data about entrepreneurs, because there were fears that these data could be used for extortion.

Ten (10) respondents also fear misinterpretation (RB2), as some data "*could result in misconceptions, influencing erroneous political decisions or the creation of incorrect services*" (I14, I15, I18, I19). However, this rarely prevents data opening and is rather a risk publishers coexist with. I1 indicates that they do not experience the barrier because they "*accompany datasets with the well-developed documentation*."

Data protection concerns (RB4) were among the most prominent issues - 13 out of 18 respondents who were able to comment on this are concerned about violating data protection legislation when sharing data. For some organizations "*this is the decisive factor to understand whether the data will be published" (I1) "because it entails unwanted responsibilities*" (I11, I12). For others, this does not prevent data publishing, but is a problem that they "*try to combat*" (I3, I14) by turning to relevant bodies (e.g., special committees) (I1-I4, I21) to discuss issues arising from unclear laws (I4, I6, I10), including "*lack of basic knowledge such as GDPR*" (I10) or lack of clarity regarding "*what anonymous data really are*" (I4). Some organizations do not find this a barrier, which they justify by the nature of the data, which do not contain personal data (I9, I18) or are provided in an anonymous or aggregated manner (I9).

Most respondents (11) reported fear of unintentionally exposing sensitive data (RB5), with the reasons largely overlapping with those associated with RB4. Organizations lacking advisory bodies or privacy expertise (including GDPR) were especially affected (I10), while those handling non-sensitive datasets reported minimal concern (I2).

**Tradition Barriers.** Responses vary regarding the reluctance of organizations to implement the cultural change required for openly sharing government data (TB2). While many organizations (15) express no strong reluctance, indicating an openness-driven approach or an established culture of publication to some extent (e.g., I2, I3, I4, I9), others highlighted limited time, training, and resources (I8, I10) or a lack of understanding of OGD's value (I16, I17). For half of respondents, skills shortages (TB3) were generally not perceived as outright barriers, although still seen as a certain challenge. More critical obstacles are attributed to (1) resource constraints (including human resources and the fear of losing skilled staff) (I1, I2) or (2) specific format requirements.

Overall, while some organizations embrace a culture of openness, many acknowledge the challenges of transitioning from traditional approaches to a more sharing-oriented culture, especially since most public sector organizations' system development traditions prioritize internal use ("*long tradition of developing in-house, and being self-contained*" (I14)), "*making it challenging to adapt these systems for broader information access and sharing*" (I13).

**Image Barriers.** Fear of published data being associated with incorrect conclusions drawn from OGD analysis by users with consequent damage to the reputation of an organization (IB5) was common and found severity, although it also does not become a decisive factor not to publish data. However, organizations dealing with more sensitive data or data on a topical issue concerning users (e.g., mobility or crisis-related data) tend to experience criticism (I3, I6). As such, I3 strongly agrees that associating with incorrect conclusions from OGD analysis could damage their reputation, citing examples of crises and increased interest to stay informed regarding their management (I15).

**Countermeasures to the Barriers.** The qualitative approach allowed participants to provide rich insights into why certain barriers were not experienced and which strategies they used to alleviate those that were. Based on these accounts, we identified 23 countermeasures that can help public agencies address the ten most widely experienced and severe barriers described in the previous section.

These countermeasures span *internal actions* (e.g., process redesign, improved data validation, enhanced documentation), *collaborative approaches* involving OGD portal providers or government bodies (e.g., publishing workflow support, infrastructure improvements, policy and compliance guidance), and *engagement with external expertise* (e.g., anonymization, quality auditing). The distribution of responsibilities highlights that overcoming OGD resistance is not the task of a single stakeholder but requires coordinated interventions across the open data ecosystem, where public agencies address internal process challenges, portal providers support publication through automation and usability improvements, government bodies shape the enabling environment through policy, legal clarity, funding, and capacity building, and external experts offer specialized capabilities that many agencies lack in-house. Figure 2 presents the identified countermeasures, mapping them to the main barriers within each category and to the stakeholder(s) responsible for implementing them.

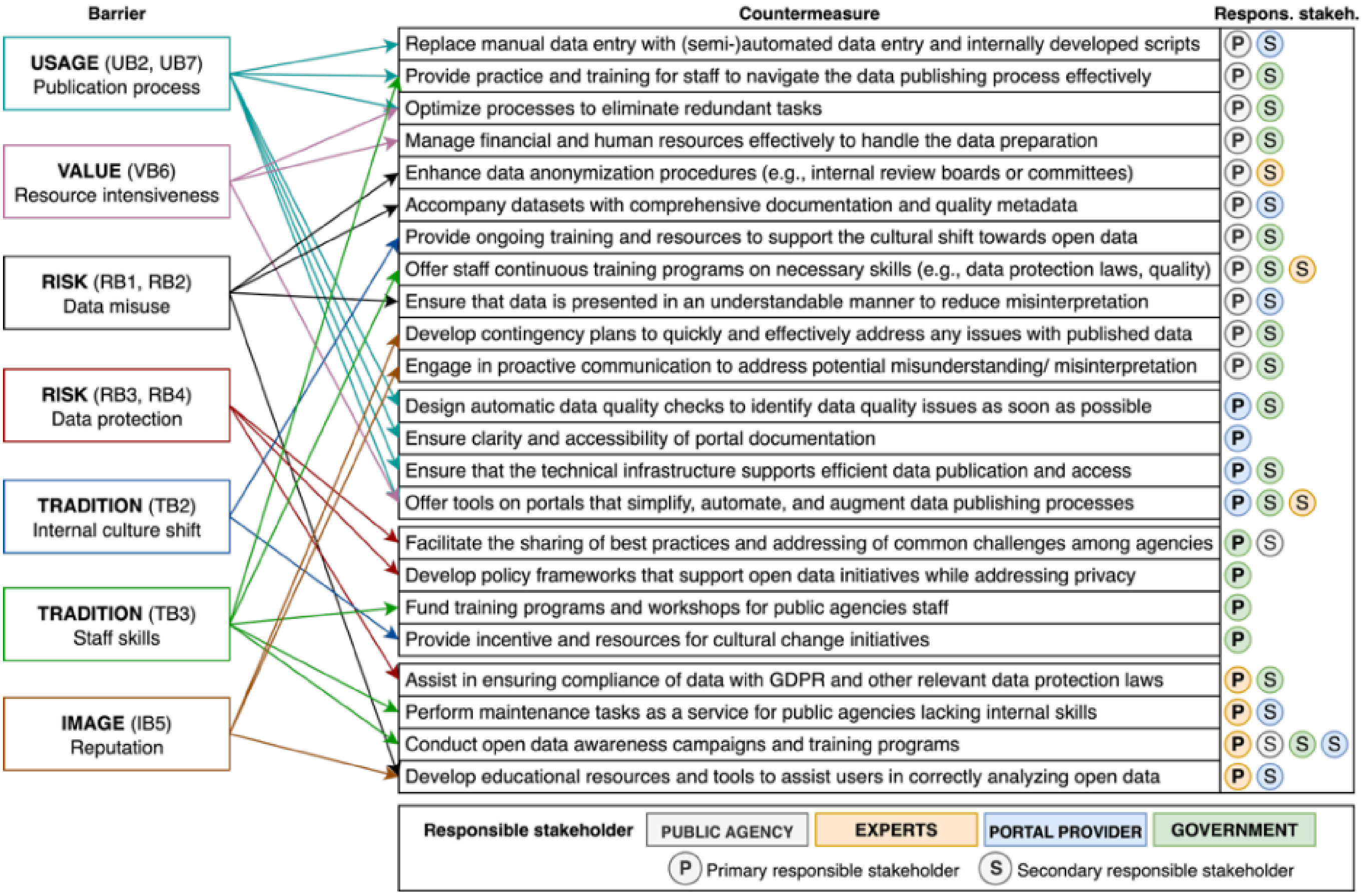

**Figure 2.** Countermeasures mapped to the most severe barriers to OGD sharing.

## 4.2 Revised IRT Model Tailored to OGD

The assessment of construct relevance showed broad agreement on the overall structure of the proposed model, with most barriers, especially within the image, tradition, value, and risk categories, considered meaningful (Figure 1). Nine barriers were viewed at least partly relevant by more than half of respondents.

**Table 3.** Final set of barriers in the IRT4DS model.

| | **Barrier** |
|---|---|
| **Usage** | UB1: It is difficult to attain the appropriate quality level for data to be shared openly<br>UB2: It is difficult to prepare data for publication so that they comply with OGD principles<br>UB3: It is difficult to prepare data for publication so that they become appropriate for reuse<br>UB4: Data are difficult to publish on the OGD portal due to the complexity of the process<br>UB5: Data are difficult to publish on the OGD portal due to the unclear (poorly defined) process<br>UB6: Data are difficult to publish on the OGD portal due to their limited functionality<br>UB7: OGD portals often do not allow for semi-automation of the publishing process<br>UB8: It is difficult to maintain openly shared government data (keep data and metadata made openly available on the OGD portal updated)<br>***UB9:*** *It is difficult to prepare, publish and/or maintain data due to lack of human resources*<br>***UB10:*** *It is difficult to prepare and publish data in required formats (i.e., those that differ from those stored in the organization, as well as in "advanced" formats, including GIS data or RDF)*<br>***UB11:*** *It is difficult to prioritize data for publication (i.e., understand what data to publish, prioritize data to be published first, as well as what data can or cannot be published)* |
| **Value** | VB1: My organization believes that openly sharing data is often not valuable for the public<br>VB2: Open government datasets are not appropriate for reuse<br>VB3: Many open government datasets suffer from data quality issues (completeness, accuracy, uniqueness, consistency etc.)<br>VB4: The public gains of openly sharing government data are often lower than the costs<br>VB5: Organizations' gains of openly sharing government data are often lower than the costs<br>VB6: Data preparation is too resource-consuming for my organization<br>VB7: Open government data do not provide any value to my organization<br>*[VB8: Open data that my organization can openly share will not provide value to users – removed from the model, it is covered as part of IB2]*<br>VB8: The amount of resources to be spent to prepare, publish and maintain open government data outweigh the benefit my organization gains from it |
| **Risk** | RB1: My organization fears the misuse of openly shared government data<br>RB2: My organization fears the misinterpretation of openly shared government data<br>RB3: My organization fears that openly shared government data will not be reused<br>RB4: My organization fears violating data protection legislation when openly sharing data<br>RB5: My organization fears that sensitive data will be exposed as a result of opening its data<br>RB6: My organization fears making mistakes when preparing data for publication<br>RB7: My organization fears that users will find existing errors in the data<br>RB8: My organization fears that openly sharing its data will reduce its gains (otherwise the organization could sell the data or use it in another beneficial way)<br>RB9: My organization fears openly sharing data will allow its competitors to benefit from it |
| **Tradition** | TB1: Freedom of information requests are sufficient for the public to obtain government data<br>TB2: My organization is reluctant to implement the culture change required for openly sharing government data<br>TB3: Employees in my organization lack the skills required for openly sharing government data<br>TB4: Employees in my organization lack the skills required for maintaining openly shared data<br>TB5: My organization is reluctant to radically change the organizational processes that would enable openly sharing government data<br>***TB6****: My organization is not actively sharing government data because data opening is not part of the culture from the very beginning* |
| **Image** | IB1: My organization has a negative image of open government data<br>IB2: My organization believes that open government data is not valuable for users<br>IB3: My organization fears that openly sharing government data will damage its reputation<br>IB4: My organization fears that the accidental publication of low-quality data will damage its reputation<br>IB5: My organization fears that associating it to incorrect conclusions drawn from OGD analysis by OGD users will damage its reputation |

Based on the relevance scores and subsequent author discussions, we refined the model to a final set of 39 items (Table 3). Several barriers were clarified to ensure clearer distinctions, particularly between initial data publication and ongoing maintenance (e.g., UB2, UB4–UB6, UB8, VB5, VB7, IB2). VB8 and IB2 were merged due to their overlap. Three additional usage barriers emerged from the interviews, namely (1) lack of human resources, (2) lack of resources to publish in the required formats, and (3) difficulty in prioritizing data (UB9–UB11). A new tradition-related barrier ("*data opening not being part of the organization's culture*") was added as TB6. Although some barriers appeared conceptually close (e.g., VB5 with VB7, UB4 with UB5), we retained them for analytical granularity, providing clearer definitions where needed. We also distinguished more explicitly between barriers affecting publication and those affecting post-publication maintenance. In cases of strong overlap, such as IB2 and VB8, we kept only the former. UB2 was retained despite initial suggestions to remove it, given divergent interpretations of OGD principles among respondents.

# 5 Discussion

This study introduces a novel theoretical model that shifts attention from factors supporting the *acceptance* of OGD [47,62] to the *resistance* that prevents public agencies from sharing their data. By applying IRT, we offer a more nuanced understanding of the interplay between functional and psychological mechanisms that shape decisions about openness. In contrast to widely used adoption theories such as TAM or UTAUT, centred on positive determinants of intention, IRT enables the systematic identification of inhibiting factors that remain obscured in acceptance-oriented perspectives. This helps explain why OGD initiatives often stall despite supportive regulatory frameworks, infrastructures, and strategic commitments.

Our findings confirm several well-documented challenges, such as data protection concerns, fear of exposing sensitive information, and the complexity of preparing data for publication [24,48,60]. However, this study identifies multiple barriers that are underexplored or absent in prior OGD literature, thereby extending theoretical understanding of why public data remains closed. These include (a) **format-specific capability gaps**, where agencies struggle to convert data into advanced or mandated formats (e.g., GIS, RDF); (b) **capacity mismatch**, reflecting the absence of dedicated OGD-specific staffing rather than general resource scarcity, pointing to structural, not merely quantitative, constraints; (c) **historical-path-dependence barrier**, where openness conflicts with inherited organizational mandates and routines, extending traditional accounts of "*risk-averse culture*"; (d) **ongoing maintenance burden** distinct from initial publication; (e) **financial cannibalization risks** for agencies relying on revenue-generating data services; (f) **difficulty prioritizing datasets** with agencies often facing uncertainty not only of *whether* but also of *which* data to release first when capacity is limited. Together, these barriers illustrate how IRT uncovers deeper organizational, operational, and institutional dynamics that traditional adoption frameworks overlook.

As a practical contribution, validating the model across 21 agencies in six European countries enables governments, open data coordinators, and policymakers to diagnose resistance more precisely and design targeted interventions. The model can support readiness assessments for national and regional data initiatives, including those arising from the European Data Act and emerging cross-sectoral data spaces. Beyond national OGD programs, the model can inform analyses in business-to-government (B2G) data sharing [51] and support efforts to strengthen local data ecosystems, where many barriers -capacity, skills, culture- are even more pronounced [50].

Complementing the diagnostic model, we identify 23 countermeasures mapped to the strongest barriers and the stakeholders responsible for acting upon them. These include **process automation**, **metadata quality practices**, **targeted training**, **cultural change initiatives**, **data protection safeguards**, and **communication strategies**. Rather than general recommendations, this mapping provides a concrete, actor-specific blueprint for addressing resistance in real OGD ecosystems and highlights that many barriers demand coordinated multi-stakeholder action, not merely technical fixes.

## 5.1 Limitations and Future Work

This study has several limitations, most of which outline directions for future research. First, the qualitative approach, though essential for developing and refining the model, limits generalizability. However, qualitative methods are well suited for uncovering context-specific mechanisms of innovation resistance [6] and align with recommendations in IRT research, which suggests qualitative inquiry as a precursor to quantitative validation [20]. Future work should test the IRT4DS model quantitatively across a larger sample of public organizations. Second, IRT is often most powerful when applied to contexts where reluctance to adopt is pronounced [23]. Although we selected countries positioned as "*beginners*" or "*followers*" in the Open Data Maturity index, it remains possible that interviewees were relatively advanced OGD practitioners within their national contexts. This could mean that some barriers were under-reported. Expanding the sample to include less mature organizations or those explicitly struggling with OGD would help capture a fuller range of resistance mechanisms.

# 6 Conclusions

This study presents an IRT-based model for OGD -IRT4DS- to explain why public agencies resist sharing their data. Drawing on literature review and interviews with 21 public organizations across six European countries, we refine the model to 39 barriers organized across five IRT categories, providing a coherent and theoretically grounded view of OGD resistance. The results confirm established challenges but also reveal several new or underexamined barriers, e.g., format-specific capability gaps, absence of OGD-dedicated staff (capacity mismatch), historical path dependence, the ongoing maintenance burden of datasets, financial cannibalization concerns, and difficulties prioritizing data for publication. These insights extend OGD research by uncovering mechanisms of resistance rarely captured by technology acceptance–oriented models and demonstrate the value of IRT for studying organizational openness. Complementing the model, we identify 23 countermeasures mapped to responsible stakeholders, illustrating that overcoming OGD resistance requires coordinated technological, organizational, and cultural interventions. Together, the model and countermeasures provide actionable guidance for practitioners and strategic insights for policymakers seeking to strengthen open data ecosystems. Overall, the IRT4DS model advances scholarship by offering a structured explanation of why public data often remains closed and by highlighting critical levers for enabling sustainable, innovation-oriented OGD initiatives.

**Disclosure of Interests.** The authors have no competing interests to declare that are relevant to the content of this article.

**GenAI use disclosure.** During the preparation of this work the authors used ChatGPT-4o in order to improve readability and language. After using this tool/service, the authors reviewed and edited the content as needed, and take full responsibility for the content of the publication.